# Microscopic phase separation in triangular-lattice quantum spin magnet κ-(BEDT-TTF)$_2$Cu$_2$(CN)$_3$ probed by muon spin relaxation

Saori Nakajima, Takao Suzuki[1,2], Yasuyuki Ishii[1], Kazuki Ohishi[1], Isao Watanabe[1],
Takayuki Goto, Akira Oosawa, Naoki Yoneyama[4,5], Norio Kobayashi[4]
Francis L. Pratt[3] and Takahiko Sasaki[4,5]

*Physics Division, Faculty of Science and Technology, Sophia University,*
*7-1 Kioicho, Chiyodaku, Tokyo 102-8554 Japan*
[1]*Advanced Meson Science Laboratory, RIKEN Nishina Center for Accelerator-Based Science,*
*Hirosawa, Wako, Saitama 351-0198, Japan*
[2]*College of Engineering, Shibaura Institute of Technology*
*307 Fukasaku, Minuma-ku, Saitama-city, Saitama, 337-8570 Japan*
[3]*ISIS Facility, STFC Rutherford Appleton Laboratory, Chilton,*
*Oxon OX11 0QX, United Kingdom*
[4]*Institute for Materials Research, Tohoku University, Sendai 980-8577, Japan*
[5]*Japan Science and Technology Agency, CREST, Tokyo 102-0075, Japan*

The ground state of the quantum spin system κ-(BEDT-TTF)$_2$Cu$_2$(CN)$_3$ in which antiferromagnetically-interacting $S$=½ spins are located on a nearly equilateral triangular lattice attracts considerable interest both from experimental and theoretical aspects, because a simple antiferromagnetic order may be inhibited because of the geometrical frustration and hence an exotic ground state is expected. Furthermore, recent two reports on the ground state of this system have made it further intriguing by showing completely controversial results; one indicates the gapless state and the other gapped. By utilizing microscopic probe of μSR, we have investigated its spin dynamics below 0.1 K, unveiling its microscopically phase separated ground state at zero field.

The ground state of the antiferromagnetic spin systems on the equilateral triangular lattice attracts considerable interest, because the geometrical frustration inhibits a simple magnetic order, so that appearance of an exotic ground state is expected[1-7]. The organic molecular charge-transfer salt κ-(BEDT-TTF)$_2$Cu$_2$(CN)$_3$, where BEDT-TTF stands for bis(ethylenedithio)-tetrathiafulvalene, is a Mott insulator and considered to be a quantum

Heisenberg antiferromagnet on the triangular lattice. The salt contains two dimerized BEDT-TTF in a unit cell, and one hole carrier localizes on each dimer due to strong Coulomb repulsive onsite correlations. Then the system can be considered as a dimer Mott insulator. An $S=½$ localized spin on each localized hole forms a nearly equilateral triangular lattice reflecting a characteristic κ-type arrangement of BEDT-TTF molecules.

So far, $^1$H/$^{13}$C-NMR[8,9] and specific heat measurements[10] have shown that there is no magnetic order at low temperatures down to 20 mK, which is much lower than the exchange interaction $J \approx 250$ K, though it maintains gapless. This result has been believed to be quite legitimate in that the frustration effect may suppress a possible magnetic order. However, quite recently M. Yamashita et al.[11] has found that the thermal conductivity shows the thermal-activation-type temperature dependence, suggesting a gapped ground state with Δ=0.46 K. This observation is calling back many arguments on its ground state until now. Furthermore, the gap survives the field of 10 T, though the thermal conductivity depends significantly on magnetic field[11]. Furthermore, NMR spectra and transverse-field (TF) μSR [13] at such high fields report the existence of field-induced static moments[9,12]. Thus, results on these experiments with different techniques bear a definite discrepancy, which requires an immediate resolution.

We have investigated the ground state of the system under zero- or extremely weak fields below a few oersted from a microscopic point of view by μSR technique, which probes a magnetic field and its dynamics inside crystals by utilizing a magnetic moment of muon. A preliminary μSR experiment had been performed sometime ago by Ohira et al.[16] who reported a slight increase in the zero-field (ZF) muon spin depolarization rate λ at low temperatures below 3 K. However, ZF-λ is dependent both on the local field amplitude $H_{local}$ and the characteristic frequency $\omega_c$ of the spin fluctuation, so that it does not identify what change in the spin state takes place. In this study therefore we worked on longitudinal decoupling (LF) measurements to detect directly the Fourier spectrum of spin fluctuation[21] in single crystals of κ-(BEDT-TTF)$_2$Cu(CN)$_3$ to understand its magnetic state.

Recently, an anomalous charge degrees of freedom is probed in this system as a dipole glass-like dielectric response at $T < 50$ K. In addition, the freezing of charge distribution fluctuation is clearly observed at $T = 6$ K [17]. The anomaly is theoretically argued to be a segregation of the charge in the half-filled HOMO onto either side of the BEDT-TTF molecules in the dimer[14,15]. At the same temperature, Manna *et al.* [18] has reported a sharp peak in the thermal expansion constant, an indicative of a second order phase transition. The origin of these anomalies at 6 K is not yet assigned its origin, that is, whether or not it concerns only lattice and charge system, because NMR-$T_1^{-1}$ also shows a steep reduction at the same temperature[8,9].

The sample of single crystals for μSR were grown by T. S. with an electrical chemical oxidation method and were characterized by the χ-$T$ measurements, the overall behavior of which is nearly identical to the previous report[8]. Crystals with the total amount of 67 mg were put on an area of 18×18 mm square so that the largest facet of each crystals corresponding to the *bc* plane is aligned, and wrapped carefully with 25 μm thick silver foil, and set to $^3$He cryostat or a dilution refrigerator to cool down to 50 mK. μSR measurements were performed at ISIS Riken-RAL Muon Fascility[12]. Fully polarized surface muon beam with the momentum of 27MeV/c was injected perpendicularly to the *bc*-plane of the sample. So as to ensure that muons stop within each crytal, we put two degradors of 50 μm-thick silver plates and a few Mylar sheets in front of the sample. Typical statistics for obtaining each depolarization curve was 20-30 Mev.

μSR probes a magnetic field and its inhomogeneity at the muon-stopping site within the sample through the time evolution of the muon spin polarization, which is measured as anisotropy in the direction of decayed positrons, which are mostly parallel with muon spins. This anisotropy is denoted as *A*, so-called an asymmetry, which in this work was analyzed by WIMDA[23]. In a case that uniform and static magnetic field exists in a sample, the time evolution of *A* shows oscillation, the periodicity of which corresponds to the Larmor frequency of a muon spin. When the field is spatially random, many different frequencies are overlapped and oscillation is smeared out so that the time evolution of *A* obeys the modified Gaussian, so called Kubo-Toyabe function.[20] In paramagnetic state, this random static field is dominated by slowly-fluctuating nuclear spins, that are considered to be quasi static. In the presence of a dynamically fluctuating field, on the other hand, one can obtain its Fourier amplitude with frequency $\omega=\gamma H_{LF}$ as a muon spin depolarization rate $\lambda$ under a static longitudinal field $H_{LF}$ applied along the muon spin direction. Thus, one must note that $H_{LF}$ dependence of $\lambda$ directly maps the Fourier spectrum of the electron spin fluctuation[21,22].

Now, we start with our result of zero-field μSR. The observed depolarization curves were almost identical to the previous report by Ohira *et al.*[16] Following their analysis, we fitted the curves with a single component function $G_{KT}(\Delta, t; H_{LF})\, e^{-\lambda t}$, where $G_{KT}$ is Kubo-Toyabe function, $\Delta$ the distribution width of the depolarization rate of the nuclear spin contribution, and $\lambda$, that of the electron spin contribution. The distribution width $\Delta$ was temperature independent to be 0.15 μs$^{-1}$ in the entire measurements, indicating that it is the nuclear contribution. We found that the observed data are well fitted to the function to reproduce the Ohira's result[16], that is, with decreasing temperature $\lambda$ starts to increase below 3 K, saturates at 0.5 K, and maintains constant down to 50 mK, suggesting that the ground state of the system is gapless. However, we must realize that one cannot tell from this observation what change in the spin state takes place, because ZF-$\lambda$ is proportional to the ratio the two parameters $H_{local}^2$ and $\omega_c$, and hence each contribution is not indissociable.

We therefore utilized LF experiments capable of probing the spin dynamics. Figure 1(a) shows typical depolarization curves in various temperatures. In $T > 3$ K, depolarization curves can be described by the same function with a single component as in ZF. The $H_{LF}$-dependences of $\lambda$ or the Fourier spectrum of the spin fluctuation obeys the Lorentz function, that is so-called Redfield function[24], as shown in Fig 2, indicating that the spins fluctuate paramagnetically. Here, the identity of the two spectra at 3 and 7 K, indicates that the spin state does not change in this temperature region.

Next, with decreasing temperature, the depolarization curves measured under fields above 3 Oe cannot be fitted with the single-component function used for analyses for ZF data. We then tried a two-component function $G_{KT}(t)(A_1 e^{-\lambda_1 t} + A_2 e^{-\lambda_2 t})$, where $\lambda_1$ and $\lambda_2$ are depolarization rates, different from one another and $A_1$ and $A_2$ are their fractions respectively, to obtain a good fit, including a small wiggling, which comes from the nuclear spin precession under the finite $H_{LF}$ and has no relation with a magnetic order. For it is unlikely that the spin state changes under such small fields, and in fact, there have not been reported any magnetic anomalies in this small field region, we can conclude that the depolarization functions in ZF should also have been fitted by this two-component function. This indicates that the system is microscopically phase separated into the two components below 300 mK. The volume fraction of each components estimated from the fit were $A_1=25(\pm 10)$ % and $A_2=75(\pm 10)$ %, which did not change with temperature down to 50 mK, and were independent of $H_{LF}$. Note that this phase separation takes place at zero field, showing a contrast to the field-induced inhomogeneity reported by NMR[9,12] and quite recently by TF-µSR[13].

Before investigating the behavior of the two depolarization rates, we refer the validity of the choice of fitting function. Other than the adopted two component function, we tried a stretched exponential function $G_{KT}(t)e^{-(\lambda t)^\beta}$, which is conventionally applied to systems with an inhomogeneity, and becomes a single component function used for ZF data in the limit of $\beta \to 1$. As shown in Fig. 1(b), it did not fit to the observed data with any value of $\beta$ indicating that our choice of the fitting function is plausible.

The obtained Fourier spectra of spin fluctuation in the $\lambda_2$ phase shown in Fig. 2 are nearly identical to those in high temperatures, indicating that this phase is an inheritance of the paramagnetic high temperature phase. We fitted these spectra with the Redfield function to obtain the characteristic frequency $\omega_c$ and the amplitude of the fluctuating local field $H_{local}$. With decreasing temperature, these two changed continuously from the paramagnetic state and showed a slight reduction at around 0.3 K, and recover again at still lower temperatures as shown in Fig. 3.

The behavior of $\lambda_1$ on the other hand is quite anomalous, that is, it increases with increasing $H_{LF}$ in a field region around 10 Oe. Such behavior has not been reported so far within the authors' knowledge except for a singlet dimer system of $KCuCl_3$. [25,26] The muon spin depolarization rate in its gapped state anomalously increases with $H_{LF}$ as reported by Higemoto *et al.*, who poses a possibility that injected muons break singlet dimers to create unpaired spins that may contribute to anomalously large $\lambda$. The anomalous increase in $\lambda$ with $H_{LF}$ is prominent only in a narrow temperature region slightly lower than the energy gap of the system, and becomes weakened at lowest temperatures [25,26]. This tendency also agrees with our observation for $\lambda_1$ phase.

From the observed experimental results, the spin state in $\kappa$-$(BEDT-TTF)_2Cu(CN)_3$ above 3 K can be summarized and compared with other experiments as follows. The single-component muon spin depolarization at high temperatures at $T > 3K$ demonstrates that the system is homogeneous and all the spins fluctuate paramagnetically. The fact that the Fourier spectra of spin fluctuation in 3 and 7 K in Fig. 2 coincide precisely leads to a conclusion that the 6 K-anomalies of thermal expansion [18] and dielectric constant [17] come entirely from the lattice and charge origin, but the change in the spin system. The steep change in NMR-$T_1$ at around 6 K is also explained in terms of the charge segregation onto either side of BEDT-TTF molecule. According to Abdel-Jawad *et al.* the amount of charge in the redistribution is as large as 10% [17], which is large enough to modify the hyperfine coupling tensor and hence $T_1$.

Next, with lowering temperature below 3 K, the appearance of two components in the muon spin depolarization function directly indicates the existence of microscopic phase separation to singlet phase ($\lambda_1$) and paramagnetic phase ($\lambda_2$). Combining this fact with other experimental results, we try to depict a possible spin state in $\kappa$-(BEDT-TTF)$_2$Cu(CN)$_3$ below 3 K. First, as stated above, the 6 K-anomaly has been reported many experiments and interpreted as the charge segregation[14,17,18], which has no direct effect to the spin state of the system, and hence to µSR depolarization functions. However, the charge segregation pushes the averaged position of each spin from the center to the side of the molecular dimer. This should bring the random spatial modulation to the effective exchange interaction $J$, transforming the nearly equilateral triangular lattice to the random lattice, which is quite likely to phase separate at lower temperatures as also posed theoretically[1415]. In the phase separated state, the singlet $\lambda_1$ phase newly appears, and the paramagnetic $\lambda_2$ phase inherits the uniform paramagnetic phase at high temperatures. From the absence of the thermodynamic phase transition down to 20 mK [9] one can deduce that paramagnetic $\lambda_2$ phase must be geometrically restricted to be nano-size, and that the interaction between those islands must be very weak, though the actual size must be determined by the diffuse scattering of neutron experiments.

Note that this picture of nano-size islands of the $\lambda_2$ phase separated by the singlet sea of the $\lambda_1$ phase also resolves the disagreement over the results of thermal conductivity and specific heat[10,11]. The specific heat exhibits gapless temperature dependence; this is because the phase $\lambda_1$ and $\lambda_2$ contribute independently to the specific heat and the latter bears the gapless temperature depenedence. On the contrary, as for the thermal conductivity, heat transport between islands of $\lambda_2$ phase is cut by the barrier of $\lambda_1$ phase surrounding around each island. This may bring the gapped-type thermal conductivity. The resurgence of $\omega_c$ below 0.1 K is also simply understand within this picture; when a possible phase transition is suppressed by the phase separation, a critical slowing down of the spin fluctuation is also stopped, and $\omega_c$ at low temperatures is considered to tends to recover the paramagnetic value.

In order to confirm our picture, we trace the temperature dependence of $\lambda_1$ and $\lambda_2$ at zero field as shown in Fig. 4. In high temperatures where the system is uniform, $\lambda_1 = \lambda_2$ coincides with $\lambda$. At 0.3 K, where the system phase separates, and $\lambda_1$ and $\lambda_2$ take values distant from one another, and $\lambda$ locates in between, indicating that what $\lambda$ describes is an average of $\lambda_1$ and $\lambda_2$. Interpolating the temperature dependence of $\lambda_1$, one notes that it coincides with $\lambda$ at 3 K, where $\lambda$ is to start to increase. This means that the increase in ZF-$\lambda$ corresponds to the occurrence of the phase separation, or equivalently the appearance of the $\lambda_1$ phase[10]. We re-analyzed the ZF-depolarization curve at 0.6K, for which LF data are absent, by fitting it with the two-component function with the volume fraction parameters fixed to those determined by the LF measurements. As shown in Fig. 4, thus-obtained $\lambda_1'$ and $\lambda_2'$ at 0.6 K agree well to the interpolated temperature dependence of $\lambda_1$ and $\lambda_2$ that were determined from LF experiments, assuring the validity of our phase separation model.

The temperature dependences of $\lambda_1$ and $\lambda_2$ are understood in a following way. The anomalous increase in $\lambda_1$ at low temperatures is owing to the muon perturbation effect to the singlet phase. On the other hand, $\lambda_2$ takes nearly a constant value down to 50 mK, because the slight reduction in $\omega_c$ at 300 mK shown in Fig. 4 is masked by a countermovement of $H_{local}$ and is invisible in ZF-value of the muon spin depolarization rate. This fact tells us again the usefulness and importance of the LF-decoupling to investigate the microscopic magnetism.

Finally, an emergence of inhomogeneous staggered moment in high an applied field reported in other NMR and TF-μSR[9,12,13] is understood in a following way. Our results indicate that the system is already phase separated in zero field, where the difference in the two phases underlies only in the spin dynamics. Then when a high magnetic field is applied, the difference will stabilized and emphasized to be observed as a static inhomogeneity[9,12,13]. So, it is considered that the present μSR results has given information complementary to those NMR and μSR in high field.

In summary, we have investigated the spin dynamics in the quantum spin system

$\kappa$-(BEDT-TTF)$_2$Cu(CN)$_3$ to reveal that the system phase separates at zero field below 3 K to the paramagnetic islands and singlet phase.  We have also proposed the possible model for the phase separation state of the paramagnetic microscopic islands surrounded by the singlet sea.

**Acknowledgements**

The authors are grateful to Dr. Seiko Kawamura-Ohira for offering her previous data. This research was partially supported by Grant-in-Aid's for Scientific Research of (B) 20340085, (C) 21540344 and on Innovative Areas "Molecular Degree of Freedom" (No. 21110518 and 21110504) of The Ministry of Education, Culture, Sports, Science, and Technology, Japan.

**References**
1. Leon Balents, Review-insight, Nature **464** (2010) 199-208.
2. Hiroshi Shintani, Hajime Tanaka, Nature Phys. **2** (2000) 200.
3. Hikaru Kawamura, Atsushi Yamamoto, and Tsuyoshi Okubo, J. Phys. Soc. Jpn, **79** (2010) 023701.
4. G. Baskaran, Phys. Rev. Lett. **63** (1989) 2524–2527.
5. Yang Qi and Subir Sachdev, Phys. Rev. **B77** (2008) 165112.
6. Tarun Grover, N. Trivedi, T. Senthil, and Patrick A. Lee, Phys. Rev. **B81** (2010) 245121.
7. Satoru Nakatsuji, Yusuke Nambu, Hiroshi Tonomura, Osamu Sakai, Seth Jonas, Collin Broholm, Hirokazu Tsunetsugu, Yiming Qiu, Yoshiteru Maeno, Science **309** (2005) 1697.
8. Y. Shimizu, K. Miyagawa, K. Kanoda, M. Maesato, and G. Saito, Phys. Rev. Lett. **91** (2003) 107001.
9. Y. Shimizu, K. Miyagawa, K. Kanoda, M. Maesato, and G. Saito, Phys. Rev. B **73** (2006) 140407R.
10. Satoshi Yamashita, Yasuhiro Nakazawa, Masaharu Oguni, Yugo Oshima, Hiroyuki Nojiri, Yasuhiro Shimizu, Kazuya Miyagawa And Kazushi Kanoda, Nature Phys. **4** (2008) 459.
11. Minoru Yamashita, Norihito Nakata, Yuichi Kasahara, Takahiko Sasaki, Naoki Yoneyama, Norio Kobayashi, Satoshi Fujimoto, Takasada Shibauchi and Yuji Matsuda, Nature Phys. **5** (2009) 44.

12. Atsushi Kawamoto, Yosuke Honma, and Ken-ichi Kumagai, Phys. Rev. **B70** (2004) 060510(R).
13. F. L. Pratt, P. J. Baker, S. J. Blundell, T. Lancaster, S. Ohira-Kawamura, C. Baines, Y. Shimizu, K. Kanoda, I. Watanabe and G. Saito, Nature **471** (2011) 612–616.
14. Chisa Hotta, Phys. Rev. B 82 (2010) 241104(R) [4 pages].
15. Makoto Naka and Sumio Ishihara, J. Phys. Soc. Jpn. **79** (2010) 063707.
16. S. Ohira, Y. Shimizu, K. Kanoda, G. Saito, J. Low Temp. Phys. **142** (2006) 153.
17. Majed Abdel-Jawad and Ichiro Terasaki, Takahiko Sasaki and Naoki Yoneyama, Norio Kobayashi, Yoshiaki Uesu, Chisa Hotta, submitted to Phys. Rev. **B82** (2010) 125119.
18. R. S. Manna, M. de Souza, A. Brühl J. A. Schlueter, and M. Lang, Phys.Rev. Lett. **104** (2010) 016403.
19. T. Matsuzaki, K. Ishida, K. Nagamine, I. Watanabe, G. H. Eaton, and W. G. Williams, Nuclear Instrumental Methods in Physics Research A**465** (2001) 365.
20. T. Saito, A. Oosawa, T. Goto, T. Suzuki and I. Watanabe, Phys. Rev. **B74** (2006) 134423.
21. Takayuki Goto, Takao Suzuki, Keishi Kanada, Takehiro Saito, Akira Oosawa, Isao Watanabe, and Hirotaka Manaka, Phys. Rev. **B78** (2008) 054422.
22. Takao Suzuki1, Fumiko Yamada, Takayuki Kawamata, Isao Watanabe, Takayuki Goto, and Hidekazu Tanaka, Phys. Rev. **B79** (2009) 104409.
23. WIMDA: a muon data analysis program for the Windows PC, F. L. Pratt, Physica B **289-290** (2000) 710.
24. C. P. Slichter, *Principles of Magnetic Resonance* (Springer-Verlag, Berlin, Heidelberg, 1990).
25. W. Higemoto, H. Tanaka, I. Watanabe and K. Nagamine, Phys. Lett. **A243** (1998) 80-84.
26. W. Higemoto, H. Tanaka, I. Watanabe, S. Ohira, A. Fukaya, K. Nagamine, Physica **B289-290** (2000) 172.

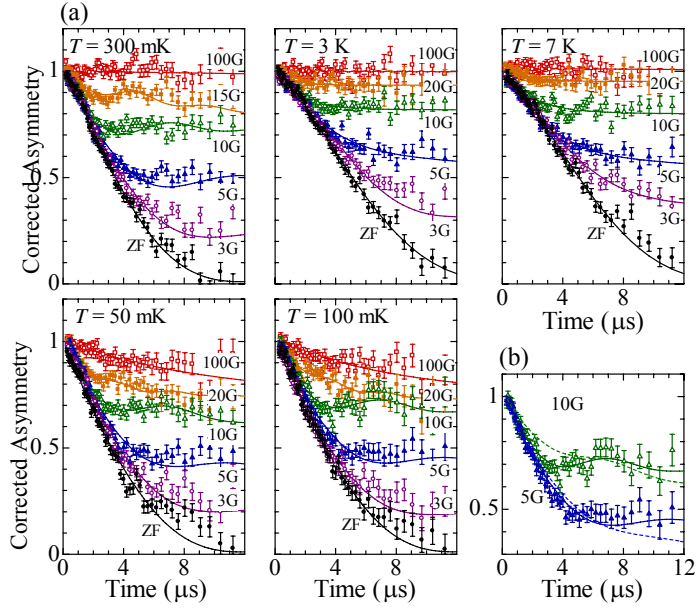

Fig 1. (a) Muon spin depolarization curves under various longitudinal fields (LF's). The curves show the fitted function (See text). (b) Fitting of observed data with the two different functions: solid and dashed curve correspond to the two-component function and a stretched exponential function.

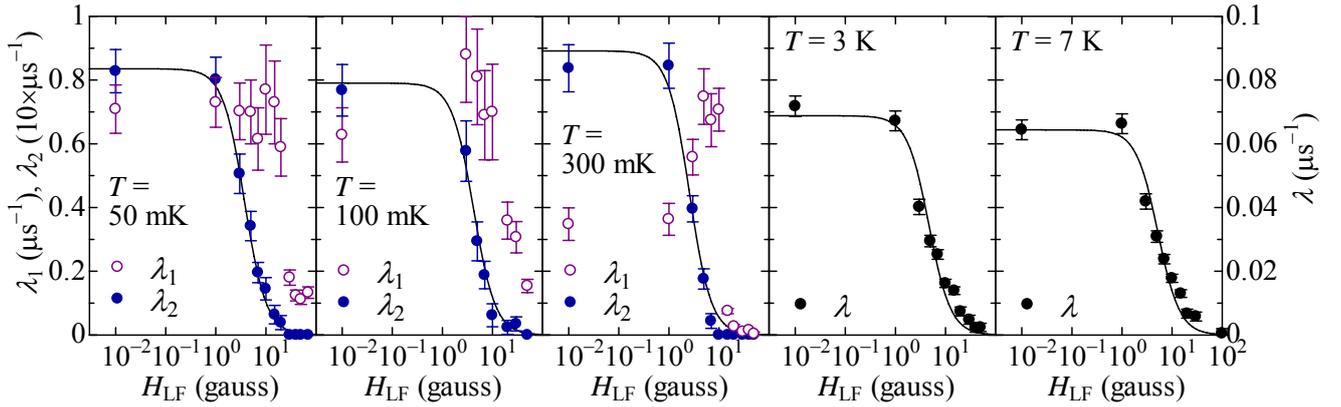

Fig 2. LF-dependence of the depolarization rate under various temperatures. Below 300 mK, the two rates $\lambda_1$ and $\lambda_2$ are separately plotted. Solid curves show the Lorentzian function, from which $\omega_c$ and $H_{local}$ are obtained. Abscissa is cut above 100 Oe, where the depolarization rate is zero within an experimental accuracy.

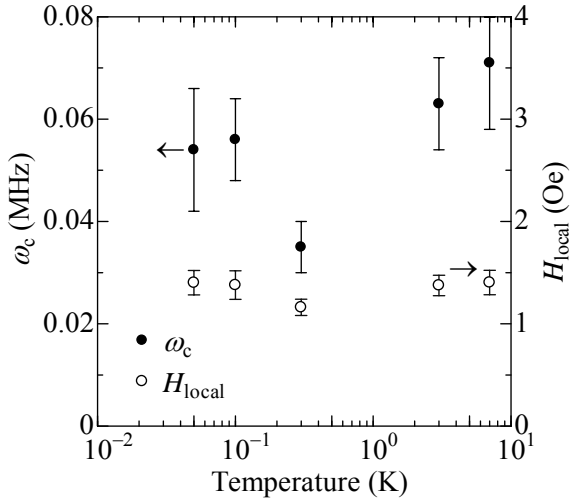

Fig 3. Temperature dependences of the characteristic frequency of the spin fluctuation $\omega_c$ and the local field amplitude $H_{local}$.

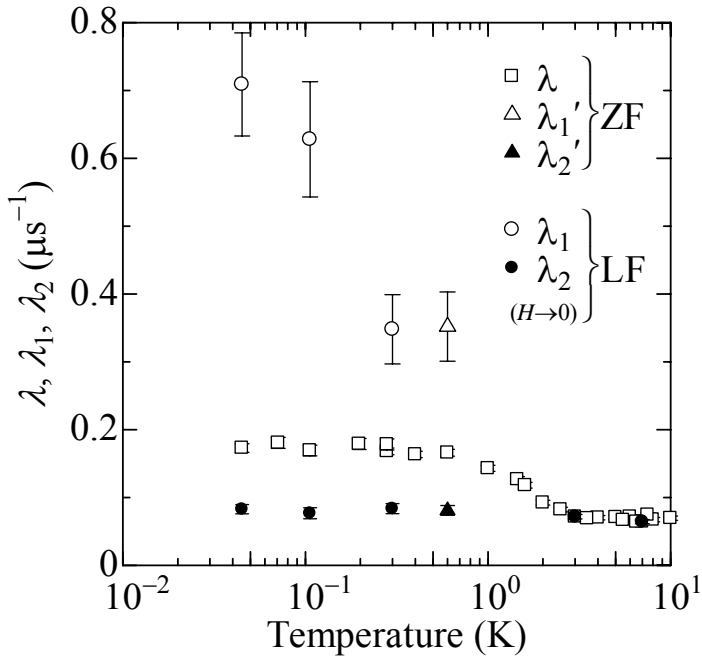

Fig 4. The temperature dependences of muon spin depolarization rates at zero field. $\lambda_1$ and $\lambda_2$ are obtained by LF measurements and extrapolated to $H=0$. $\lambda_1'$ and $\lambda_2'$ are obtained by reanalyzing the ZF-depolarization curve with the two-components fitting function. $\lambda$ is obtained by analyzing ZF-depolarization curves with the single-component function.